\documentclass{article}

\usepackage[utf8]{inputenc}
\usepackage[justification=centering]{subfig}
\usepackage[table,xcdraw]{xcolor}
\usepackage{amsmath,amssymb,amsthm}
\usepackage{algorithm,algorithmic,setspace}
\usepackage{pdfpages,multirow,enumitem}
\usepackage{graphicx,url,hyperref}
\usepackage{fancyvrb,fancyhdr,makecell}
\usepackage{multirow,array,bm,longtable}

\newcolumntype{P}[1]{>{\arraybackslash}p{#1}}
\newcolumntype{M}[1]{>{\centering\arraybackslash}m{#1}}
\begin{document}
	
	\title{Collaborative Drug Discovery: Inference-level Data Protection Perspective}
	
	\author{Balazs Pejo$^{*}$, Mina Remeli$^{**}$, Adam Arany$^{***}$, \\
	Mathieu Galtier$^{****}$, Gergely Acs$^{*****}$}
	
	\date{$^{*}$CrySyS Lab, BME, Hungary, pejo@crysys.hu (Corresponding author)\\
		$^{**}$ University of Cambridge, United Kingdom, mincsek@gmail.com (Work was done while at CrySyS Lab)\\
		$^{***}$Stadius, KUL, Belgium, adam.arany@kuleuven.be\\
		$^{****}$Owkin, France, mathieu.galtier@owkin.com\\
		$^{*****}$CrySyS Lab, BME, Hungary, acs@crysys.hu\\}
	
	\maketitle
	
	\begin{abstract}
		Pharmaceutical industry can better leverage its data assets to virtualize drug discovery through a collaborative machine learning platform. On the other hand, there are non-negligible risks stemming from the unintended leakage of participants' training data, hence, it is essential for such a platform to be secure and privacy-preserving. 
		This paper describes a privacy risk assessment for collaborative modeling in the preclinical phase of drug discovery to accelerate the selection of promising drug candidates. 
		After a short taxonomy of state-of-the-art inference attacks we adopt and customize several to the underlying scenario. Finally we describe and experiments with a handful of relevant privacy protection techniques to mitigate such attacks.
	\end{abstract}
	
	\section{Introduction}
	
	Drug discovery and development is currently a time-consuming and expensive process. Many small molecules fail in the later stages of development due to toxicity or a lack of sufficient efficacy. More complex mechanisms of action and stricter regulation have resulted in a drastic drop in pharmaceutical R\&D productivity over the past 60 years. Indeed, only 11\% of the drug candidates that enter clinical phases make it to approval, i.e.~are demonstrated to be productive and safe. As a result, it takes on average 1.9 billion euros and often more than ten years to bring one successful drug to the market. 
	
	Machine Learning (ML) is a promising approach to reducing the time and cost of this development process, where a carefully trained model can predict the biological activities of small molecules. The available volume of training data mainly determines the quality of these models. Consequently, the industry is making the first steps towards federated ML approaches that leverage more data than a single partner (e.g., a pharmaceutical company). Federated learning (FL) enables several partners to train a common model without directly sharing their training data, still obtaining a model that can predict biological activities almost as accurately as a model trained centrally on the union of their training data. Such collaborations between pharmaceutical companies can save them millions of dollars today and save lives tomorrow. Hence, there is a great interest in such collaborations both within and outside the pharmaceutical industry. Indeed, a similar scenario is currently being studied by several pharmaceutical companies within the MELLODDY project (EU/EFPIA IMI 831472).
	
	However, FL is not a panacea. Several prior works have demonstrated that FL per se does not provide sufficient protection against the unintended leakage of training data. Even if participants only exchange gradients during training and can only access the common model at the end, both the gradients and the model itself leak information about the training data. For example, as we show, exchanged model parameters can reveal that a specific participant has a particular compound in its training data, which could indicate what drugs the participant is experimenting with. Such inference attacks can seriously undermine the ultimate goal of FL and cause substantial financial loss to a company investing millions of euros into curating high-quality training data.

	In this paper, we perform a data protection risk analysis of a FL on an real use case, where several pharmaceutical companies collaboratively train a multi-task federated machine learning model to predict activities between chemical compounds (i.e., drug candidates) and different targets (e.g., proteins).
	Such federated structure-activity modeling in Drug Discovery during pre-competitive collaboration is a well-suited problem for our analysis as it captures and illustrates well the core trade-off that the privacy-aware participants face.
	In this context, data protection refers to protecting the confidentiality of intellectual properties (IPs) such as a participant's training data or some model (hyper)-parameters\footnote{Data protection and privacy are often used interchangeably in the literature, though ``privacy'' usually refers to personal data which is not the case in our application.}. Attacks aim to infer confidential attributes of these assets (such as the membership of specific training samples) and are therefore called \emph{inference attacks}. The pharma partners specify the exact attributes, and their successful inference is represented as a \emph{risk} in our methodology. We design inference attacks (threats) to estimate the likelihood of these risks. In particular, the feasibility and the success probability of these attacks measure the likelihood of the risk, which then, combined with the risk's severity (negative impact), allows for identifying the most dangerous risks to be mitigated. We are only concerned with passive inference attacks in this analysis. Indeed, perturbing the data results in a more considerable accuracy drop for a malicious party than for the others \cite{pejo1together}. Hence, there is no incentive for active attacks (e.g., poisoning, back-doors, etc. \cite{goldblum2020data}) as long as the adversary also needs good model quality. The specific contributions of this study are three-fold.
	
	%There are couple of ways how confidentiality can be defined, for instance as inference or attribution. 
	%Note that attribution can be achieved always with sufficient background knowledge. Consequently in this paper we focus on inference instead of studying various assumption about the adversary's knowledge for attribution. 
	%Moreover, for practical reasons we only consider passive inference attacks. 
	
	%Such collaborations between pharmaceutical companies can save them millions of dollars today as well as save lives tomorrow. Hence, there is a great interest in such collaborations both within and outside the pharmaceutical industry. On the other hand, besides the clear benefits there are potentially dangerous aspects of such a collaboration. The data federated models might be trained on are highly sensitive and confidential, corresponding to secret intellectual properties and a billion dollar business. Indeed, a similar scenario is currently studied by several pharmaceutical companies within MELLODDY project (EU/EFPIA IMI 831472).
	
	\begin{itemize}
		\item We demonstrate the process of data protection risk analysis on a real machine learning application. We propose a risk analysis methodology that helps systematically map different threats and hence identify risks that should be mitigated. Although we focus on risks stemming from the leakage of intellectual properties in FL, our methodology equally applies to the case when ML models process personal data and therefore are subject to data protection regulations (e.g., GDPR). Our approach is in line with the risk-based approach proposed by several regulations on AI  (such as the upcoming European AI Regulation\footnote{\url{https://digital-strategy.ec.europa.eu/en/policies/regulatory-framework-ai}}) and hence can be useful to audit ML models in practice. 
		%Following a risk analyst mindset, we identify the potential risk sources and assets which correspond to sensitive information. We identify several risks (i.e., when a risk source obtains a protected asset) and present various possible threats realizing these risks. 
		\item We design and implement two novel and scalable (membership) inference attacks. They detect information leakage through gradients of any training sample and the trained model. Even if secure aggregation is in place  \cite{bonawitz_practical_2017}, which prevents the attribution of membership information to a specific partner, we show that differential attacks can still be used to assign membership information. We develop a novel differential attack that exploits that some change in the composition of the training participants also changes the membership attack's output distribution, which statistical tests can detect. These attacks show that --- just as in other applications of FL --- gradients and the activation values of the commonly trained model can leak sensitive information about the participants' training data. 
		
		\item We propose technical countermeasures to mitigate these threats. We show that secure aggregation \cite{bonawitz_practical_2017} does prevent trivial attribution of the inferred attribute to a specific partner. Still, it is not sufficient to avoid attribution if the composition of the partners changes during training. Although Differential Privacy (DP) \cite{desfontaines2020sok} defeats such attribution, we show that DP implies intolerable accuracy drop-in practice. Hence, we explore ad-hoc mitigation approaches and test data protection empirically with our attacks. Although these techniques provide no formal privacy guarantee, they still mitigate the attacks with acceptable utility loss. For example, we find that thresholding model parameters effectively mitigate information leakage without degrading model quality too much. This also shows that model parameters are highly redundant, which allows a model to memorize complete training samples that our attacks can detect. 
	\end{itemize}
	
	%We structure the paper as follows. 
	%Section \ref{sec:pre} contains the federated learning architecture with secure aggregation which we considered for our drug discovery use-case. 
	%Section \ref{sec:risk} walks the reader through a general risk analysis process and the specific steps concerning drug discovery.  
	%In Section \ref{sec:implemented_attacks} we detail the attacks which we developed and 
	%in Section \ref{sec:def} we show the results of the proposed mitigation techniques. 
	%Finally in Section \ref{sec:conc} we conclude our paper. 
	
	\section{Related work}
	\label{sec:ex_att}
	
	We review the literature concerning inference attacks.
	There are many privacy-related attacks concerning machine learning; here, we give a non-comprehensive list of categories based on the attacker's goal. For more comprehensive surveys, we refer the reader to \cite{liu2021machine,rigaki2020survey}. Model inversion attacks aim to reconstruct a representative value of a class \cite{fredrikson_model_2015}, e.g., a record that is similar to all records belonging to a class. Membership inference attacks are aimed at inferring if a certain record was part of the target model's training dataset \cite{hu2021membership}. The most common techniques to achieve this utilize shadow models \cite{nasr_comprehensive_2018,salem_ml-leaks:_2018,shokri_membership_2016,truex_towards_2018} or rely on overfitting \cite{leino_stolen_2019,murakonda_ultimate_2019,pyrgelis_knock_2017,sablayrolles_white-box_2019}. Reconstruction attacks take membership inference attacks another step forward by identifying the whole training dataset (instead of one record) \cite{hitaj_deep_2017,li_quantification_2019,salem_updates-leak:_2019,wang_beyond_2018,zhu_deep_2019}. Property inference attacks aim to infer properties of training data that are independent of the features that characterize the classes of the joint model \cite{ganju_property_2018,melis_exploiting_2018,pejo2020good,wang_eavesdrop_2019}. And finally, model extraction attacks arise when an adversary obtains black-box access to some target model and attempts to learn a model that closely approximates or even matches the original model \cite{TramerZJRR16}. 
	
	Despite the wide range of attacks that exists, in this paper, we focus on \emph{membership inference attacks}. These are the most elemental attacks, highlighting one bit of information leakage. If membership inference succeeds, that flags (one-bit) information leakage, a requirement for other attacks, which potentially leak more bits. On the other hand, if it does not succeed, that can be a solid empirical argument that other attacks (that leak more information) will probably fail as well.
	
	Several mitigation techniques have been proposed against the risks of membership inference attacks, such as regularization or hyperparameter tuning (including model adaptation), which reduce overfitting \cite{yeom_privacy_2017}. DP \cite{desfontaines2020sok}  can also be applied to ML models providing provable privacy guarantees albeit with significant accuracy degradation in general. 
	Two dedicated defenses, adversarial regularization \cite{nasr_comprehensive_2018}, and MemGuard \cite{JiaSBZG19} have also been proposed to combat membership inference attacks. As MemGuard requires post-processing the output of the ML model, while adversarial regularization is scalable only to a minimal number of attacks, we focus on hyperparameter tuning, regularization, and DP in this work. 
	
	\section{Context: Collaborative Drug Discovery}\label{sec:pre}
	
	%This section details the utilized artificial neural network model for collaborative drug discovery. 
	
	Quantitative Structure-Activity Relationship (QSAR) models are commonly used to predict bioactivity\footnote{A beneficial or adverse effect of a (small molecular) drug on a living organism. In this context, the result of a measurement carried out on some preclinical model systems (enzyme assay, cellular assay, etc.) can indicate a potential beneficial or adverse effect of the given drug candidate. Therefore a valuable asset for the company.} of chemical compounds on different targets \cite{dudek2006computational}. The input to such models is a representation of a chemical compound (e.g., an extended-connectivity fingerprint (ECFP) \cite{rogers2010extended}) which encodes different properties of the molecule. The output is the predicted bioactivity of the input molecule on specific targets (e.g., proteins). As building such models requires substantial data, collaborative learning is a compelling approach for multiple pharma companies to benefit from each other's confidential data collectively. 
	
	Chemical fingerprints like the ECFP encodes a chemical structure in the form of a vector of binary values or counts. A typical implementation is that substructures of the molecules are hashed to an integer address pointing to a position in a high dimensional vector, and the given position is either set to one or incremented. This high-dimensional vector is usually folded to smaller sizes to decrease the model size and improve generalization. 
	
\begin{figure}[b!]
		\centering
		\includegraphics[scale=0.6]{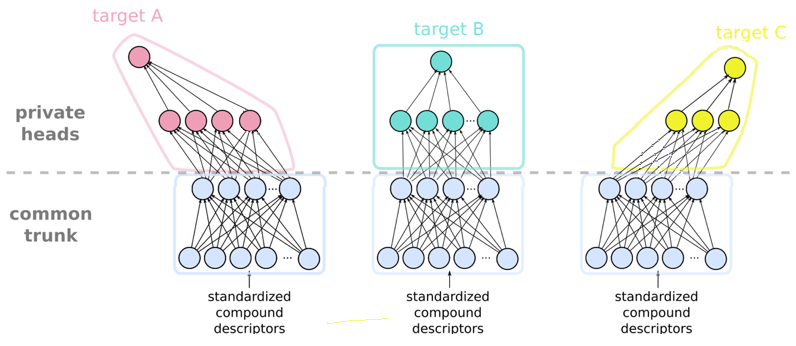}
		\caption{Head-trunk architectures with feed-forward neural networks.}
		\label{fig:headtrunk}
	\end{figure}

	This paper employs feed-forward neural networks as QSAR models and trains in a multi-partner, multi-task setting. Neural networks are highly modular and scalable, making them, especially appealing in this context. Our neural network is separated into a \emph{trunk} network and different \emph{head} networks, as illustrated in Figure \ref{fig:headtrunk}. The trunk is common for all tasks and participants and trained collaboratively, while the head architecture depends on the task and can be different among participants. The trunk extracts those features from a compound that can be used to maximize the overall prediction accuracy concerning all tasks.
	
	%\subfloat[Aggregation in collaborative multi-task learning.]{\includegraphics[scale=0.3]{HeadTrunkComm}}
	%Collaborative learning with a head-trunk architecture. Only the trunk is learnt collaboratively, whereas the training data and the architecture as well as the parameters (weights) of the head are never disclosed. The aggregator sums the model updates of every participant and re-distributes the result to each participant. Every participant updates the trunk with the same aggregate which means that a new trunk will be used in the next round. 
	
	In a collaborative learning setting, each participant trains its model (head and trunk together) on its training data locally. Only the parameter updates of the trunk are sent to the server for aggregation in every federated round. The weights and the whole architecture of the head networks are confidential and therefore not shared. The single role of the server (aggregator) is to sum up, the trunk updates sent by every participant and re-distribute the result. Each participant updates its trunk model with the common aggregate and obtains a new trunk model that is identical for all participants. 
	
	Therefore, the training data of participant $j$ consists of pairs of $(\mathbf{x}, \mathbf{y}^j)$, where $\mathbf{x} \in \{0,1\}^n$ denotes the fingerprint of the input compound  ($n$ is the folded fingerprint size and is identical for all participants), and $\mathbf{y}^j \in \mathbb{R}^{k_j}$ denotes the bioactivity values of this compound ($k_j$ is the number of targets in participant $j$'s data), that is, $\mathbf{y}_i^j$ denotes the bioactivity value between compound $\mathbf{x}$ and target $i$ for participant $j$, and $k_j$ is the output size of participant $j$'s head model. 
	
	A typical trunk architecture consists of a single linear layer. For instance, if the trunk has 6000 units and a folded input size is 32000, each participant will send and receive a total of 192 million parameter values (as model update/aggregate) to/from the server in every federated round. In each round, every participant performs a single gradient descent iteration: first, it selects a batch of training samples uniformly at random, then computes the gradients (i.e., model update), and finally sends the update to the server for aggregation. This is also illustrated in the middle (white area) of Figure \ref{fig:illustrate}.
	
	Notice that this is a typical cross-silo FL setting: collaborative drug discovery is usually executed by pharmaceutical companies equipped with the necessary training data. Hence, the number of participants is below a dozen, and each of them participates in every round (as all of them have dedicated hardware with constant availability). By contrast, in cross-device FL, only a subset of participants participate (out of possible thousands) with often intermittent connection. Besides, each participant performs only a single local gradient iteration on its training data in cross-silo FL. In contrast, participants perform multiple iterations in cross-device to increase bandwidth efficiency (at the cost of model accuracy and possibly convergence speed).
	
	\section{Methodology and Threat Model}\label{sec:risk}
	
	As none of the available regulations on AI and data privacy mandate any specific risk analysis methodology, we propose a sufficiently general and simple one to demonstrate the risk-based approach promoted by most legal requirements. 
	
	The risk analysis is composed of the following main steps repeated until all Risks become acceptable: 1) define the perimeter of the analysis, i.e., the actors (Risk Sources) and the data involved (Assets), 2) define the controls already in place, 3) define for each Risk Source the associated Risks, 4) define Threats for each Risk, and 5) apply controls to mitigate Risks. Within this paper, we do one iteration of these steps focusing on confidentiality. We rigorously go through the assets we consider, the possible corresponding Risks and their Sources, and detail the related feasibility and likelihoods. We mention a couple of Threats that are realized with our proposed attacks. Finally, we suggest several mitigation strategies with a trade-off parameter between model accuracy and defensive efficiency.
	
	%Instead, this paper focuses on evaluating the feasibility of technical attacks mounted by actors in the operational phase and not in the design phase. Conceptually, this boils down to two attackers: internal pharmas and the aggregator server. 
	
	\subsection{Data and Actors}
	
	\paragraph{Assets. }
	Assets are the confidential or private data that the Risk Sources aim to learn or manipulate, thereby causing Risk. We define the following assets: \emph{Chemical Fingerprint (FP)} (the input of the model, a compact representation of a chemical compound), \emph{Targets} (protein or others like toxicity for which the Bioactivity is being measured), \emph{Bioactivity} (the predictive task labels for the chemical fingerprint input), and the \emph{Trunk \& Head model} (the collaboratively and privately trained neural network, respectively). 
	%To understand the Risk of privacy attacks, the assets that could be compromised were defined first. Two possible assets could leak to the attacker:  the \textbf{data} used for training and the trained \textbf{model} itself.
	
	We assume that a fingerprint can be inverted, and the chemical compound represented by the fingerprint can be unambiguously identified. %Due to arbitrary hash functions used to create these fingerprints, it was widely considered non-invertible; however, this is proven wrong. 
	Indeed, in a recent work, it was shown that the chemical structures corresponding to an ECFP fingerprint could be recovered with reasonable accuracy by learning a feed-forward network to project to the hidden space of a recurrent autoencoder \cite{le2020neuraldecipher}. %In particular, if an attacker has access to a descriptor vector of a molecule (like the Morgan fingerprints) and knows how to create such a vector for a given molecule, he could use generative models with reinforcement learning (or any other generative process like genetic algorithms) driven by a measure of similarity to discover the underlying structure (Smiles) of the molecule with a high probability of success. 
	
	\paragraph{Risk Sources. }
	A Risk Source is a person or non-human source that can cause a Risk, accidentally or deliberately. We only focus on human sources (aka, attackers), which aim to partially or entirely infer some confidential properties of the assets. 
	We define two primary Risk Sources: \emph{Pharma}, who is represented as a participant in collaborative learning, and \emph{Server}, who coordinates and aggregates the training. 
	This paper is NOT concerned with Risks and Threats stemming from outsider attacks, including but not limited to a cloud provider, external pharma companies, hackers, etc. Attackers are characterized by their background knowledge and their access to assets.
	
	%Attackers can be categorized along many axes, e.g., Active or Passive, depending on whether the attack requires modifying the messages, algorithms, or training data, or it only performs the attack without deviating from the learning protocol.
	
	\paragraph{Background knowledge. }
	Based on the attacker's role in the collaborative learning setting, they can access different data. The attack may need information about the training data, and depending on its extent, there are many categories. For example, the attacker could access the training dataset with or without labels or know only about the statistical properties (e.g., its distribution). 
	
	\paragraph{Access. }
	The attacker can access all the details of the model (white-box) or only its prediction (black-box). In the drug discovery use case, this can be further specified due to the multi-task environment, e.g., in case of a white-box attack, does the attack access the weights and architecture of the commonly shared trunk model or the private head? The attacks could be separated based on the available meta-information, too: the attacker might monitor and use the performance metrics or utilize communication traffic patterns during collaboration, such as timing, size, or direction). Finally, the frequency of the exploitable information could also play a significant role in the attack's success. 
	
	\subsection{Utilized Control: Secure Aggregation}
	\label{sec:secagg}
	
	Secure aggregation \cite{AcsC11,bonawitz_practical_2017} is a standard technique to prevent access to the model update of an individual partner, and it allows the server (and the participants) to learn only the aggregated model updates as illustrated in the middle (white area) of Figure \ref{fig:illustrate}. In other words, nothing is leaked except the aggregated model in every training round. Hence, inference (i.e., attribution) of any confidential information to any participant is not possible from the \emph{aggregated} update, if all participants are involved in the aggregation in every round. 
	
	This is similar to secure multi-party computation (SMC) \cite{cramer2015secure}, where the intuitive security requirement is that nobody can learn information about the participants (e.g., dataset, gradients, etc.) than what they could already know from the aggregated result and their own input. Rather than for the entire training process, secure aggregation implements an SMC protocol in each round separately. Indeed, besides the final model, the intermediate (secure) aggregates are also revealed in secure aggregation. This is a standard compromise in the literature \cite{bonawitz_practical_2017} as end-to-end SMC techniques are still quite challenging for complex models with many parameters and large datasets \cite{KMA+19} due to performance issues, which is precisely the case with drug discovery.
	
	Before sending the model update (gradients) of the Trunk to the servers, the participants \textit{mask} their update in such a way that the masks cancel out once aggregated (by the server). Hence, the server only learns the sum of model updates and not those of the individual members of this sum. Therefore, the server can only learn information about the set of participants as a whole without being able to attribute the information to any specific participant (e.g., it might be able to infer that a certain compound is used during the federated run, but it cannot assign this compound to any specific participant).
	
	%For an aggregator, who does not contribute to the computation with any input, it means that the aggregated model (or sum) should be the only learnt information from the protocol messages. %A typical example for such federated learning scenario is depicted in Figure \ref{fig:fed.learning.scenario}.
	
	%\begin{figure}[h]
	%    \begin{center}
		%    \includegraphics[scale=0.1]{sec-agg-scenario.png}
		%    \end{center}
	%    \caption{The federated learning framework, according to \cite{SGA20}. At iteration $t$, the central server sends the current version of the global model, $x(t)$, to the contributors of the federated learning (e.g., mobile users, pharma companies, etc.). User $i\in[N]$ updates the global model using its local data, and computes a local model $x_i(t)$. The local models are then aggregated in a privacy-preserving manner. Using the aggregated models, the central server updates the global model $x(t + 1)$ for the next round, and pushes it back to the users.}\label{fig:fed.learning.scenario}
	%\end{figure}
	
	%Secure Aggregation \cite{bonawitz_practical_2017} is an indispensible protocol to prevent trivial attribution attacks even when all parties participate in the training and their composition does not change over time. Indeed, without secure aggregation, a malicious party or server could access individual model updates and utilize several of the previously mentioned inference attacks to gain knowledge that can be linked to a specific partner. 
	
	%We assume the Trunk gradients are aggregated with the secure aggregation protocol. 
	Note that secure aggregation does not eliminate privacy breaches completely: it does not hide the number of trunk parameters, and hence may leak information about the trunk's architecture to any Risk Source (e.g., the server) who can access the encrypted model updates.
	%Second, a sufficiently knowledgeable adversary might always be able to link sensitive  information to a specific participant.
	Furthermore, the aggregated model could also be concealed from the server: we assume each participant can add an extra mask which is known to every participant except the aggregator. After aggregation this mask can be removed by any participant, i.e., the mask persists and do NOT cancel out after aggregation.
	
	\subsection{Risks}
	
	%\begin{figure}[h]
	%    \centering
	%    \includegraphics[width=10cm]{summary.pdf}
	%    \caption{Summary of Risks. Circles and boxes denote assets and the main group of Risks, respectively.}
	%\end{figure}
	
	%\red{Update Figure: remove Ghosting/Poisoning, update Risks IDs}
	
	A Risk represents the goal of the Risk Source which is to infer confidential information about one of the asset. Its motivation is characterized by the incentives (e.g., financial gains) of the Risk Source to perpetrate the Risk balanced with the potential disincentives (loss of trust, damage to reputation, etc.). A Risk's impact is measured by its associated Severity. In this paper, we only consider inference Risks, consequently, neither poisoning (intentional contamination of the training process) \cite{Backdoor_Fed} nor free riding (participation in the training process with intentionally limited contribution) \cite{hardin2003free} is enlisted as a Risk in Table \ref{tab:fearedevents}. We also do not consider all identified assets, our goal is not to provide a comprehensive list of risks but to demonstrate the methodology on a real-world machine learning use-case. 
	
	\begin{table}[b!]
		\centering
		\begin{small}
			\begin{tabular}{|M{.04\linewidth}|M{0.5\linewidth}|M{0.1\linewidth}|M{.1\linewidth}|}
				\hline
				\textbf{ID} & \textbf{Description of Risk} & \textbf{Risk Source} & \textbf{Severity} \\ 
				\hline
				\hline 
				R1 & \makecell{A fingerprint is inferred with \\  high probability to have been utilized by \\ A) a given participant B) any participant.} & Pharma Server & 4 \\ 
				\hline
				R2 & \makecell{Targets -- which was not published before -- \\can be inferred with high probability to \\ have been utilized by \\ A) a given participant B) any participant.} & Pharma Server & 4  \\
				\hline
				R3 & \makecell{(A part of) the trunk model is \\ inferred/reconstructed.} & Server  & 3  \\
				\hline
			\end{tabular}
		\end{small}
		\caption{Identified Risks. Severity:  1 – Negligible, 2 – Limited, 3 – Significant, 4 – Maximum. Only R1 is considered in this paper.}
		\label{tab:fearedevents}
	\end{table}
	
	\subsection{Threats}
	
	A Threat is a sequence of actions carried out by a Risk Source to realize one or more Risks. A Threat is sometimes called an ``attack'' in the security literature. In general, several Threats can lead to a Risk and the same Threat can lead to different Risks. A Threat has a single property called feasibility, which characterizes the difficulty for a Risk Source to realize the Threat. The likelihood of a Risk is computed from the feasibility of its corresponding Threats and is influenced by the feasibility as well as the the motivation or risk source. To ease demonstration, we assume in the sequel that 1) the likelihood of the Risk is computed directly from the feasibility of its Threats using standard probability calculus (see Section \ref{sec:def}), and 2) only a few threats corresponding to risk R1 and listed in Table \ref{tab:threaths} are considered. We are not concerned with Threats that manipulate the learning algorithm in the design phase by model architects or software designers (they are also excluded from the list of potential risk sources).
	
	\begin{table}[t!]
		\centering
		\begin{tabular}{|M{.03\linewidth}|M{.45\linewidth}|M{.08\linewidth}|M{.06\linewidth}|M{.09\linewidth}|}
			\hline
			\textbf{ID} & \textbf{Description of Threat} & \textbf{Feas. \& Like.} & \textbf{Corr. Risk} & \textbf{Risk Source} \\
			\hline
			\hline
			T1 & \textbf{Membership attack} against the compound: for a given set of compounds, an attacker having access to the trunk model can compute their representation, and build a classifier which tells if they are in the training set.  & 2 & R1B & Server  \\
			\hline 
			T2 & \textbf{Membership attack} against the compound: based on the model updates, an attacker builds a classifier which tells if a compound is in the training set. & 4 &  R1B & Server  \\ 
			\hline
			T3 & Combination of T1 with traffic metadata collection to assign identified compounds to certain partners.  & 1 &  R1A & Pharma Server  \\ 
			\hline
		\end{tabular}
		\caption{Identified Threats for Risk R1 from Table \ref{tab:fearedevents}. Feasibility/Likelihood: 1 - Very Difficult/Negligible, 2 - Difficult/Low, 3 - Average/Moderate, 4 - Easy/Large, 5 - Immediate/Certain. }
		\label{tab:threaths}
	\end{table}
	
	%\paragraph{Traffic analysis}
	
	%Although all data are encrypted with secure aggregation and a TLS channel is additionally established between each participant and the aggregator, the meta-data can still potentially be used to infer sensitive information about pharmas. For example, partners may be located in different continents and time-zones and hence traffic pattern such as packet timing, routes, direction, traffic size, can all make a partner's datastream unique which may facilitate attribution attack. Indeed, several attacks in the literature \cite{CaiZJJ12} have exploited such meta-data to launch successful de-anonymization attacks against the communicating parties.
	
	%De-anonymization attacks seem impossible in the presented setup, as long as all security primitives are implemented correctly, and each participant sends data for aggregation in every single round. In particular, secure aggregation requires each partner to send data with identical format (i.e., update vectors with the same size), otherwise data vectors cannot be summed. As long as there is no other communication between the parties beyond secure aggregation, and data is being transferred in the same format from every single partner in every single round, then traffic analysis is not possible. %Clearly, \emph{robust} secure aggregation can break this guarantee, since failing parties may send data individually.
	
	\section{Attacks}
	\label{sec:implemented_attacks}
	
	\begin{table}[t!]
		\begin{tabular}{|M{0.16\textwidth}||M{0.22\textwidth}|M{0.1\textwidth}|M{0.36\textwidth}|}
			\hline
			\textbf{Attack}                  & \textbf{Leaked Asset}                       & \textbf{Corr. Threat} & \textbf{Adversarial Capabilities}         \\ \hline\hline
			\textbf{Trunk Activation Attack} & training data                               & T1      & Access to final trained trunk model. \\ \hline
			\textbf{NGMA}                    & training data                               & T2      & Access to aggregated trunk gradients in a single round.                 \\ \hline
			\textbf{N-1 Attack}              & assigning training data to a single partner & T3      & Access to aggregated trunk gradients in multiple rounds.    \\ \hline
		\end{tabular}
		\caption{Summary of presented attacks.}
		\label{tab:attacks_summary}
	\end{table}
	
	In this section, three different attacks are described. One can be performed after (or during) training (Trunk Activation), while two (NGMA and $N-1$) are applicable only during training. The motivation and background of the attack, the attacker's capabilities, and the attack itself are described for each attack. A short summary is depicted in Table \ref{tab:attacks_summary} and an illustration is visualized in Figure \ref{fig:illustrate}. 
	
	\subsection{Attack settings}\label{sec:set}
	
	\begin{table}[b!]
		\centering
		\begin{footnotesize}
			\begin{tabular}{|l|c|c|c|c|c|c|c|c|c|c|}
				\hline
				Samples & 256 & 432 & 4301 & 7076 & 5576 & 32954 & 35692 & 45796 & 64903 & 68359 \\
				\hline
				Tasks & 2 & 8 & 22 & 46 & 32 & 182 & 150 & 217 & 309 & 340 \\ 
				\hline
			\end{tabular}
		\end{footnotesize}
		\caption{The meta-data of the ChEMBL dataset split utilized in our experiments.}
		\label{tab:split}
	\end{table}
	
	All evaluations are performed on the publicly available ChEMBL v.23 dataset \cite{gaulton_chembl_2017}, prepared according to \cite{simm2019graph}. The input to this model is a folded ECFP fingerprint of a compound with a size of $n= 32 000$, and the output is the predicted bioactivity with 3547 classification tasks (out of which 1308 were used for simulating the ten participants). The trunk has 40 units, all with ReLU activation and a 0.2 dropout. We used the SparseChem library\footnote{\url{https://github.com/melloddy/SparseChem}}. The implementation of the proposed attacks is open-sourced on GitHub\footnote{\url{https://github.com/minaremeli/pharma-attack-framework}}.
	
	Cross-silo FL is simulated with a realistic separation of the training data into ten partitions, each with different samples and tasks (i.e., output size). The metadata of the splitting is shown in Table \ref{tab:split}.
	
	\subsection{Membership Inference}
	
\begin{figure}
		\centering
		\includegraphics[width=\textwidth]{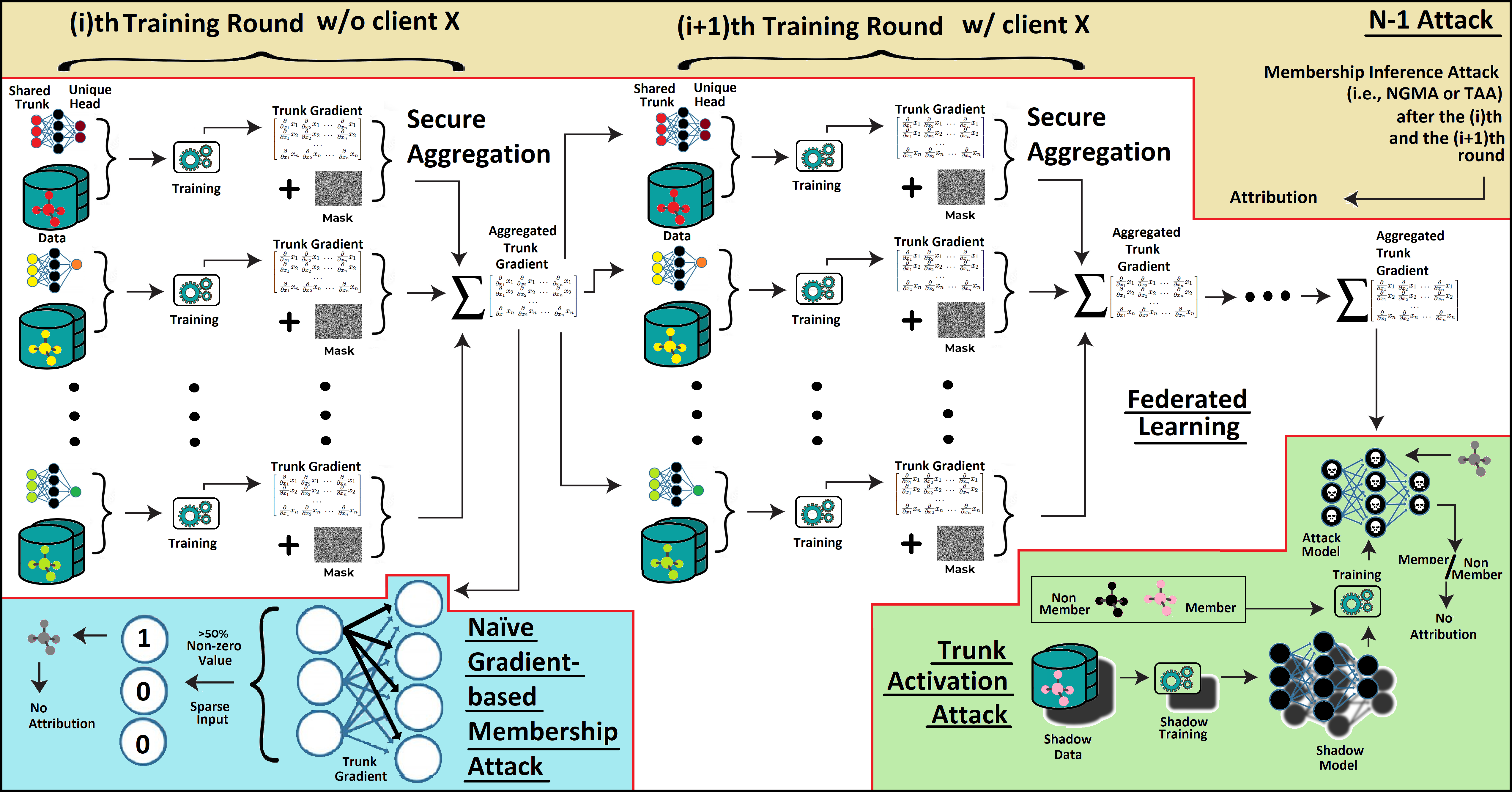}
		\caption{Illustrating the proposed attacks and the process with secure aggregation (the middle white area). The attacks are from Table \ref{tab:attacks_summary}, i.e., Trunk Activation Attack, Naive Gradient-Based Membership Inference Attack, and the $N-1$ Attack illustrated on the bottom right light green, the bottom left light blue, and the top light khaki area respectively. Each attack is detailed in the section they are introduced. }
		\label{fig:illustrate}
	\end{figure}

	The trunk is shared among all participants, who might be honest but curious. Therefore, it is interesting to understand whether its output (i.e., trunk activation values) or its updates (i.e., gradients) leak any information about the membership of an input sample. Moreover, the commonly trained trunk model may be sold after the federated run to third parties, posing additional risks. Next, we show two membership attacks: a Trunk Activation Attack and a Gradient-based Membership Attack, where the former infers membership from the output of the trunk model after (or during) training. At the same time, the latter does the same from the model update during training.
	Recall that (as illustrated in Figure \ref{fig:illustrate}) the attacker cannot access individual model updates due to secure aggregation; hence the output of the attacks only indicates whether any participant used the tested sample for training or not without attributing this information to any specific participant.
	
	\subsubsection{Trunk Activation Attack}
	
	The hypothesis is that the trunk will produce different outputs (activation values) for a training sample compared to a yet unseen sample. An attacker model is built to detect these subtle differences between training and non-training samples. We assume that the attacker knows the target model architecture (white-box attack). 
	
	The attack has three phases (also illustrated on the bottom right (light green area) of Figure \ref{fig:illustrate}): the shadow model training phase, the attack model training phase, and the evaluation phase, where the attacker model is evaluated on the targeted model. In the first phase, the attacker builds a shadow model that can simulate the targeted model's activation output. (S)he trains it with his shadow dataset, which comes from the same distribution as the participants' data. In the second phase, (s)he trains the attack model on the activations of the shadow model and learns to differentiate the activation patterns of members from that of non-members, e.g., between the black compound and the pink one where only the latter is part of the shadow dataset. The attack model is a RandomForest classifier with 100 estimators and no maximum depth. The output of the attacker model is binary, indicating whether any of the participants used the sample to train the trunk. Finally, the last step is to query the RandomForest classifier with the activations of samples; thus, the attacker model can infer the membership of sample (compound) that the attacker has access to (and not only of a particular sample), without attributing that to any specific client.
	
	\paragraph{Evaluation. }
	The attacker aims to build a classifier that distinguishes trunk activations of samples that are part of the training dataset and samples which are not. For this, it needs to obtain trunk activations of both training and non-training sample, which will constitute the training data of the attacker model. For this purpose, it builds its copy of the original model. The attacker's dataset is divided into two disjoint parts $D$ and $D'$. $D$ is used for the federated training of the trunk model, and $D'$ contains samples that the trunk model has never seen in the training process. The trunk's activation values are used to train the attacker model (i.e., a random forest classifier): samples from $D$ are selected and labeled as ``seen", while samples from $D'$ are labeled as ``unseen". $D$ and $D'$ together are used to train and validate the attacker model (a random 66\% of the data is used for training and the remaining for validation). In particular, for every $(x,y) \in D'$ (analogously to $D$) the trunk activation values for compound $x$ is computed. This results in a vector of 40 activation values, which are fed to the attacker classifier to predict the label $y \in \{seen, unseen\}$.
	
	\paragraph{Results and Feasibility. }
	The mean accuracy of the attack is \textbf{61.98\%}, with a precision of 60.78\% and a recall of 67.59\%. Consider the random guess as a baseline, where the attacker has a chance of 50\% to guess the correct membership without any prior knowledge. Therefore, because this attack needs labeled data to train the attack model, the feasibility of this attack is estimated to be ``\emph{Difficult}'' for a Pharma partner and ``\emph{Very difficult}'' for the Server who cannot access the trunk model (see Section \ref{sec:secagg}). 
	
	% \begin{table}[h!]
		%\centering
		%\begin{tabular}{|
		%>{\columncolor[HTML]{9FC5E8}}l |
		%>{\columncolor[HTML]{9FC5E8}}l |l|l|}
	%\hline
	%\multicolumn{2}{|l|}{\cellcolor[HTML]{9FC5E8}}                                                         & \multicolumn{2}{l|}{\cellcolor[HTML]{CFE2F3}\textbf{actual values}} \\ \cline{3-4} 
	%\multicolumn{2}{|l|}{\multirow{-2}{*}{\cellcolor[HTML]{9FC5E8}\textbf{}}}                              & \cellcolor[HTML]{CFE2F3}positive & \cellcolor[HTML]{CFE2F3}negative \\ \hline
	%\cellcolor[HTML]{CFE2F3}                                            & \cellcolor[HTML]{CFE2F3}positive & {\color[HTML]{3D3C40} 12922}     & {\color[HTML]{3D3C40} 3694}      \\ \cline{2-4} 
	%\multirow{-2}{*}{\cellcolor[HTML]{CFE2F3}\textbf{predicted values}} & \cellcolor[HTML]{CFE2F3}negative & {\color[HTML]{3D3C40} 6823}      & {\color[HTML]{3D3C40} 15876}     \\ \hline
	%\end{tabular}
	%\vspace{0.1cm}\\
	%\begin{tabular}{|l|l|l|}
	%\hline
	%\rowcolor[HTML]{CFE2F3} 
	%\textbf{Mean accuracy} & \textbf{Mean precision} & \textbf{Mean recall} \\ \hline
	%73.52\%                & 66.04\%                 & 77.88\%              \\ \hline
	%\end{tabular}
	%\caption{The confusion matrix of the trunk activation attack and the evaluation of the trunk activation attack on 442672-long inputs.}
	%\label{tab:TAA_eval}
	%\end{table}
	
	\subsubsection{Naïve Gradient-based Membership Attack (NGMA)}
	\label{sec:ngma}
	
	A simple membership attack, called Naïve Gradient-based Membership Attack (NGMA) exploits the uniqueness and sparsity of the ECFP fingerprint (as illustrated on the bottom left (light blue area) of Figure \ref{fig:illustrate}): on average there are less than 70 values in the 32k long folded fingerprint in the ChEMBL dataset. This uniqueness is also reflected in the updates (gradients) of the trunk (noted with darker and lighter arrows).
	
	In particular, if some coordinate values of the input fingerprint are zero, then all the gradients corresponding to these zero coordinates will be zero as well. When the gradient is computed on a batch with multiple samples, some parts of the aggregated gradients may become zero as the gradients of different samples can potentially cancel each other out. Because this can happen quite often, we require that out of all the weight updates (gradients) connected to a single non-zero input neuron, the majority of them must be non-zero. If all non-zero input neurons of a compound ''pass'' this test, the compound is likely to be involved in the corresponding training batch, i.e., an unknown participant(s) used it.
	
	In details, let $S$ denote the non-zero positions in the fingerprint of the tested sample, and let $\frac{\partial L}{\partial w_{i,j}}$ denote the gradient belonging to weight $w_{i,j}$ in the trunk ($L$ is the loss function and $w_{i,j}$ are the weights). If $\frac{\partial L}{\partial w_{i,j}}\not=0$ for all $i\in S$ and more than half of all $j$ in the update, then the fingerprint is deemed to be used in the current training batch.
	
	NGMA is much more feasible and scalable than building a separate attacker model to classify model updates, which is also pursued by prior works \cite{nasr_comprehensive_2018}. Indeed, it does not require training any attacker model, and therefore the attacker does not need any training data beyond the tested fingerprint. Moreover, NGMA does not require knowledge of the label of the tested samples.
	
	%\begin{figure}[h!]
	%    \centering
	%    \includegraphics[scale=0.5]{input_layer_gradient.pdf}
	%    \caption{Gradients ``connected'' to non-zero inputs are non-zero themselves.}
	%    \label{fig:NGMA_fig}
	%\end{figure}
	
	% Let $S$ denote the non-zero positions in the fingerprint of the sample of interest. Furthermore, let $\frac{\partial L}{\partial w_{i,j}}$ denote the gradient belonging to weight $w_{i,j}$ in the trunk ($L$ is the loss function). If $\frac{\partial L}{\partial w_{i,j}}\not=0$ for all $i\in S$ and more than half of all $j$ in the update, then the fingerprint was used in the current training batch. 
	% \red{Do we need $\partial$ to explain this? Cannot we do it in high level as the rest of the paper?}
	
	%The pseudo-code of the attack is as follows:
	
	%\begin{Verbatim}[frame=single]
	%target_in_batch = True
	%for each nonzero input_n:
	%    positive_votes = 0
	%	for each neuron_m:
	%		if (GRADIENTS[neuron_m, input_n] > 0)
	%			positive_votes += 1
	%	if (positive_votes < M/2)
	%        target_in_batch = False
	%\end{Verbatim}
	
	\paragraph{Evaluation. }
	% The federated learning setting is simulated with 9 participants. Each participant has 26242 training samples (compounds with 2808 activation values), and in each round, a batch of 524 samples is processed per participant (this means that the global batch has $9\cdot524=4716$ samples). 
	Using NGMA, the attacker (i.e., the server or any participant) aims to infer whether a particular (target) compound is used as a training sample. For one experiment, 50 positive and 50 negative batches are collected and evaluated; a positive/negative batch includes/excludes the target (one batch is composed of the union of all participants' batches). Model updates on the 50 positive and 50 negative batches were computed, and the attack was then executed on each of them. This evaluation process was repeated 200 times, each with a random target. %The confusion matrix with average TP, FP, TN, and FN values is shown in Table \ref{tab:NGMA_eval}.
	
	\paragraph{Results and Feasibility. } 
	NGMA can determine with an average accuracy of \textbf{82.41\%} whether a specific compound was in the current batch. The precision and recall are 74.17\% and 100\%, respectively. NGMA is more accurate than the Trunk Activation attack. Therefore its feasibility is ``\emph{Large}`` to a Pharma partner who can access the gradients, but almost negligible for the server who cannot access them (neither of their sum) due to the employed secure aggregation protocol. Recall that, unlike Trunk Activation Attack, the attacker does not need to train attacker models for NGMA and hence does not need labeled training data, which increases its feasibility.

\subsection{N-1 (differentiation) attack}
\label{sec:n1attack}

In cross-silo FL, parties may join or leave during training as illustrated on the top (light khaki area) of Figure \ref{fig:illustrate}. Therefore, using a successful membership test, the attacker can attribute a training sample to the leaving/joining party. For example, if the membership test starts to succeed upon the arrival of a new party (e.g., client $X$), then the party is likely to have the tested sample in its training data. Similarly, when the membership test starts to fail upon a party's leaving, the party is likely to have the tested sample in its training data. Such an attribution attack is a differentiation attack, as it takes advantage of the change in the coalition. The $N-1$ attack is a particular case of the differentiation attacks when the difference is a single partner. However, other differentiation attacks are also possible: for instance, in the cross-silo FL scenario, where only a (random) subset of participants participate in each round, attribution can become possible by accumulating the information of a membership attack from each round \cite{pejo2020good}.

The following evaluation aims to measure how successful an ``$N-1$'' attack can be on ChEMBL.
A simulated scenario is where participants train together for some time; then, one randomly selected participant leaves. The target sample (on which the membership inference attack is performed) is part of the dataset of the leaving party. The attacker performs a membership inference attack ($MI$) against the target sample in each training round and records the result of the attack; for each round $r$ and each target sample $T$, $MI(r, T)$ output whether or not the target is present in that training round. These are called \textit{positive} and \textit{negative} rounds, respectively. In general, when the target sample is present in only one of the datasets, the attack signals that the target is present once every epoch (an epoch in the training process consists of multiple rounds). According to the attack, epochs that do (not) contain the target sample are called positive (negative) epochs.
The aim is to infer whether the statistics before are significantly different from those after the party leaves using Fisher's exact test.
%Here, the aim is to infer whether the attacker can confidently deduce that the target sample belongs to the leaving party using a binomial test. 
%In particular, the occurrence count of positive/negative epochs can be described with a binomial random variable with two outcomes. In the training phase, where all the participants train together, the attacker records the frequency of the positive outcomes. However, when the membership inference attack fails once the participant with the target sample leaves, the frequency of the positive epochs is expected to drop. If the frequencies in the two cases can be reliably distinguished with a binomial test, then the attack is considered successful.

\paragraph{Evaluation. }
We assume that the number of positive epochs is much higher than after leaving before the party leaves. The attacker runs the membership inference in every round and, using the attack outputs, determines whether the epoch was positive or negative. First, the attacker does this when everyone trains together (for 30 epochs). Then, after the party leaves, he observes another 30 epochs. The null hypothesis is that the number of positive epochs should be the same in the first 30 epochs as in the second 30 epochs. 

\paragraph{Results and Feasibility.}
In the case of 30 epochs, 26 were flagged as positive when the particular participant participated in the training, and 0 epochs were flagged as positive after they left.
With the Fisher-test, we calculate the probability of this data, given our null hypothesis. The probability of observing 26 positive epochs, then 0 is precisely $3.92 \cdot 10^{-13}$. This is also equal to the significance level of a one-tailed test, which tells us the probability of precisely getting this (or more extreme) data (with the same marginals).
%Both of these correspond to $p$-value 0.00, which is smaller than the 5\% significance threshold; therefore, the null hypothesis is rejected. This means that the frequency of the positive epochs significantly changes when a party leaves. 
Therefore, the attacker can confidently conclude that the target sample was part of the leaving party's dataset since the number of positive epochs decreased significantly. This means that the attribution of a tested compound to a specific party is not less feasible than the membership attack itself, no matter whether secure aggregation is employed. That is, \emph{secure aggregation does not mitigate the risk of attribution  (R1A and R2A) as long as the composition of the parties changes during training.} 

\subsection{Discussion}

To compute the likelihood of the Risk R1 in Table \ref{tab:fearedevents}, consider the feasibility of the two attacks; NGMA with feasibility ``Easy`` (i.e., an attack accuracy of 0.82), and TrunkActivation with feasibility ``Difficult`` (i.e., an attack accuracy of 0.6) for a Pharma partner as a Risk Source. Risk 1B (membership inference without attribution) occurs if at least NGMA or TrunkActivation succeeds, which has a probability of $1-(1-0.82)(1-0.6)\approx 0.93$\footnote{We assume they are independent events. } 
%$E_i$, $\Pr[\vee_i E_i] = 1 - Pr[\wedge_i \overline{E_i}]$. 
This amounts to a substantial likelihood with maximum severity; hence, this risk must be mitigated. The same holds for Risk 1A since attribution is very feasible with our ``$N-1$ attack'' described in Section \ref{sec:n1attack}.  

\section{Mitigation}\label{sec:def}

In this section, several techniques (controls) are described to mitigate the threats in Table \ref{tab:attacks_summary} and eventually Risk 1 in Table \ref{tab:fearedevents}. An overview of these techniques and the mitigated attacks (i.e., Threats) are summarized in Table \ref{tab:mitigations}. %In conclusion, a handful of compression techniques to decrease secure aggregation's computation and communication costs are also presented. 
In general, there are legal, organizational, and technical controls. Legal controls include any contractual agreements or legal obligations that the parties must obey. Organizational controls include training employees, establishing working policies, and re-structuring the organization to support effective data protection.    
Technical controls employ technical measures such as access control, cryptographic techniques, sanitization algorithms, detection algorithms, etc. This document focuses only on technical controls, excluding IT security countermeasures.

\begin{table}[b!]
\centering
\begin{tabular}{|c|c|c|}
	\hline
	\textbf{Mitigation} & \textbf{Addressed Threats} & \textbf{Effect on Acc.}\\ \hline \hline
	Differential Privacy & T1, T2, T3 & Severe \\ 
	\hline
	Regularization & T1, T2, T3 & Mild \\ 
	\hline
	Parameter Tuning & T1, T2, T3 & Mild \\ 
	\hline
	Compression  & T1, T2, T3 & Mild \\ 
	\hline
\end{tabular}
\caption{Mitigation techniques with the corresponding threats and accuracy distortion effect. }
\label{tab:mitigations}
\end{table}

\subsection{Differential Privacy}

Privacy (or confidentiality) could be measured with Differential Privacy (DP) \cite{desfontaines2020sok}, and in particular with parameters $\varepsilon$ and $\delta$. DP provides a strong guarantee that the membership (i.e., the presence or the absence) of a single data point will not change the final output of the algorithm significantly. DP is usually achieved by perturbation, e.g., the participants add Gaussian noise to their model updates before transferring them to the aggregator. Besides providing privacy protection, this could have a regularization effect and hence could boost generalization \cite{DworkFHPRR15}; however, excessive noise can also degrade model performance (i.e., slower convergence and lower model accuracy). DP has two flavours in the context of collaborative learning; Record-level \cite{abadi2016deep} and Client-level DP \cite{client-DP-ETH-Zurich,client-DP-McMahan}. Client-level DP guarantees that no participant-specific information about the training data of any participant is shared (up to $\varepsilon$ and $\delta$). At the same time, Record-level DP prevents information disclosure specific to any single record (up to $\varepsilon$ and $\delta$). 
%For example, Record-level DP allows us to infer that a group of compounds may have large activity with a certain target except if this group is composed of only a single compound (again up to $\varepsilon$ and $\delta$). By contrast, Client-level DP hides group information if it can be observed only within a single participant's training data. 
Although the Client-level DP guarantee is stronger, it also needs larger perturbation and provides weaker utility. To the best of our knowledge, Client-level DP is the only solution that provides a proven defense against $N-1$ attacks.

\paragraph{Evaluation. }
Since DP does provide a theoretical privacy guarantee against membership inference, we do not evaluate the protection efficiency against the presented membership inference attacks. Instead, we focus on their utility loss and evaluate model quality with centralized learning while guaranteeing Record-level DP. We show that model performance is unacceptable even in this simplified setting and, therefore, would be even worse in collaborative learning with Client-level DP. Indeed, Client-level DP requires larger perturbation than Record-level DP, and collaborative learning cannot have better performance than centralized learning on the same training data (i.e., the union of the participant's training data). 

\begin{figure}[b!]
\centering
\includegraphics[width=\linewidth]{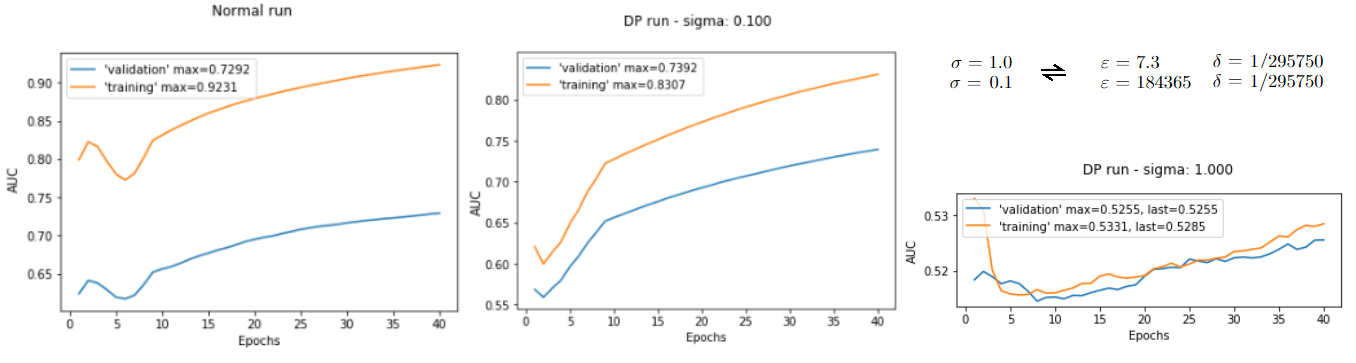}
\caption{Experimental results using Differential Privacy}
\label{fig:DP}
\end{figure}

Privacy protection is measured via the privacy parameter $\varepsilon$ and $\delta$, which can be directly mapped to the variance $\sigma$ of the Gaussian noise added to the gradients of the model parameters in every SGD iteration of the training. Contrary to the multi-partner setting described in Section \ref{sec:set}, we simulate a centralized learning scheme by using the preprocessed ChEMBL dataset from \cite{simm2019graph} without splitting. This contains 295750 samples and 2808 targets, so $\delta$ is set to $1/295750$ (reciprocal of the number of original training samples). The gradients are clipped to have an L2-norm less than $2.5$, and $\varepsilon$ is computed using the moments accountant method \cite{abadi2016deep}.

Results are shown in Figure \ref{fig:DP}. It is visible that adding a small noise level (i.e., $\sigma=0.1$) has some regularization effect (i.e., decreases the difference between the training and validation AUC values). Hence, it does not only improve privacy guarantees but also helps generalization. On the other hand, the corresponding privacy guarantee is weak ($\varepsilon=184365$). To have a meaningful privacy guarantee, one must apply noise with greater magnitude (i.e., $\sigma=1.0$, which corresponds to $\varepsilon=7.3$). However, in such a case, the utility of the training is completely eradicated. 

\subsection{Other Privacy-Preserving Methods}

The above experiments show that adding DP noise to the gradients without client-level protection leads to significant accuracy loss. Fortunately, some techniques provide ``ad-hoc'' privacy protection with tolerable accuracy loss at the same time. Although these techniques do not offer rigorous privacy guarantees, they will be shown to mitigate the inference attacks described in Section \ref{sec:implemented_attacks} reasonably well. 

\paragraph{Regularization.} 
As a regularization technique, dropout is already valuable for the neural network: it improves generalization and prevents overfitting. The effectiveness of membership inference attacks is partially due to the overfitting nature of ML models. Therefore, any techniques that increase the generalizability of the model help fight against membership attacks. 

\begin{figure}[b!]
\centering
\subfloat[Effect of dropout in input layer on NGMA attack.]{\includegraphics[scale=0.3]{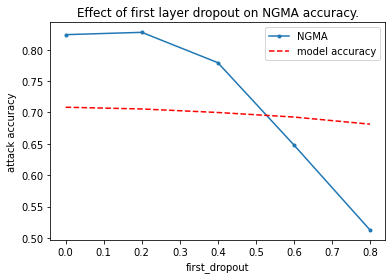}}
\subfloat[The effect of batch size on NGMA. With increasing batch size the attack accuracy drops.]{\includegraphics[scale=0.3]{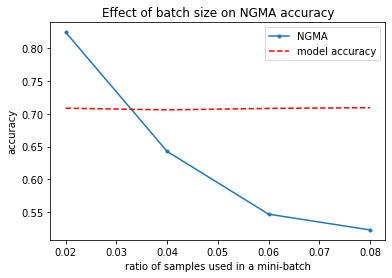}}
\subfloat[The effect of Top-K compression on NGMA and Trunk Activation Attack.]{\includegraphics[scale=0.3]{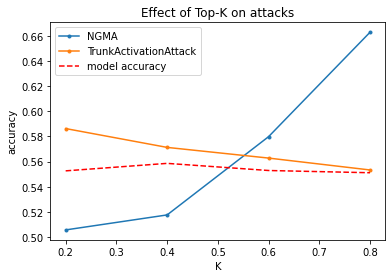}}
\caption{Various mitigation techniques without formal privacy guarantees. The blue line denotes the attacker model's accuracy after 1000 rounds of training while the red dashed line is the attacked model's accuracy. }
\label{fig:mit_NGMA}
\end{figure}

Concerning NGMA (in Section \ref{sec:ngma}) we experimented with various dropouts. We observed that while a dropout of 0.2 does not significantly impact the attack accuracy, a value of 0.6 and above does. The left part of Figure \ref{fig:mit_NGMA} shows that, with increasing dropout, there is a near-linear drop in the accuracy of NGMA. In contrast, model accuracy slightly deteriorates even for larger dropout values.

% Similar experiments were conducted with changing the activation function in the target model: with Tanh activation instead of ReLU, the attack accuracy dropped from 74\% to 60\% (while maintaining a near-100\% recall). This suggest that some model architecture choices are more robust against membership inference attacks than others, a phenomenon also highlighted in \cite{papernot2020tempered}.

%Another mitigation technique is gradient clipping, which is also utilized in relation with Differential Privacy. Although no guarantee is provided without the added noise, sheer gradient clipping without noise could decrease the privacy attack's success rate. For instance, it can  protect against backdoor attacks \cite{sun2019can}, however, in general it does not provide robustness \cite{menon2019can}.

\paragraph{Hyper-parameter tuning. }
Most prior works on membership attacks, and likewise our attack in Section \ref{sec:implemented_attacks}, considered collaborative learning when 
each participant performs a single local SGD iteration with a single mini-batch to compute its update. 
%However, we also observed that the attack accuracy deteriorates by increasing the number of mini-batches. On the other hand, increasing the number of mini-batches is not desirable as it makes convergence slower which may yield poorer model quality. 
Increasing the size of the mini-batch deteriorates the attack accuracy, as shown in the middle of Figure \ref{fig:mit_NGMA}. As the gradients of every sample in a batch are averaged in SGD, a single compound has a smaller influence on the model update if the batch is large, thus making their detection in a membership test more difficult. Therefore, increasing the batch size and the number of local SGD iterations can mitigate inference attacks at the cost of minor utility degradation. 

%\paragraph{Traffic shaping}

%The goal of traffic shaping is to prevent traffic analysis. In particular, even if traffic is encrypted, certain meta-information (timing of packets, direction of communication, size of packets) can still help infer private information about a participant (perhaps combining with other inference attacks) \cite{CaiZJJ12}. For example, if there is only one participant from a certain time-zone then the timing of packets can reveal when it contributes to the model. Traffic shaping changes the traffic pattern so that traffic analysis becomes too costly for an attacker.
%\emph{Traffic shaping mitigates privacy threats at the cost of additional communication overhead}.

%\paragraph{Shuffling of fingerprints}

%To prevent the inversion of fingerprints, the original fingerprint, which is a binary vector, can be shuffled before uploading it to the federated learning platform. Each pharma is assumed to share a secret key to perform the same  shuffling on the same fingerprint as well as to invert this shuffling. This key is shared by all data contributing partners and thus not with the aggregator. The transformation must be applied to each new input that is fed to the model (for training and prediction).
%By hiding the process of generating the descriptor vector for a given molecule the attacker can not use the measure of similarity to guide a generative algorithm.

\begin{table}[b!]
\centering
\begin{tabular}{|c|c|c|c|c|}
	\hline
	Technique & Parameter & Accuracy & Trunk Act. Acc. & NGMA Acc.\\
	\hline\hline
	No Compression &  & 70.84\% & 61.98\% & 82.41\%\\
	\hline
	Thresholding & 0.001 & 73.42\% & 54.02\% & 50.00\%\\
	\hline
	Top-K &  \begin{tabular}{@{}c@{}}
		0.2  \\ 0.4 \\ 0.6 \\ 0.8
	\end{tabular}  & \begin{tabular}{@{}c@{}}
		55.14\%  \\ 55.84\% \\ 55.27\% \\ 55.1\%
	\end{tabular}  &  \begin{tabular}{@{}c@{}}
		58.62\%  \\ 57.13\% \\ 56.28\% \\ 55.33\%
	\end{tabular} &  \begin{tabular}{@{}c@{}}
		50.57\%  \\ 51.76\% \\ 57.98\% \\ 66.28\%
	\end{tabular}\\
	\hline
	Random Subset & \begin{tabular}{@{}c@{}}
		0.2  \\ 0.4 \\ 0.6 \\ 0.8
	\end{tabular}  & \begin{tabular}{@{}c@{}}
		54.31\%  \\ 53.54\% \\ 55.29\% \\ 55.54\%
	\end{tabular} &  \begin{tabular}{@{}c@{}}
		60.42\%  \\ 57.44\% \\ 61.04\% \\ 59.52\%
	\end{tabular} &  \begin{tabular}{@{}c@{}}
		50\%  \\ 50\% \\ 50.02\% \\ 59.54\%
	\end{tabular}\\
	%         \hline
	%         Quantization & 16 & 52.09\% & 58.18\% & 50.00\%\\
	\hline
\end{tabular}
\caption{The effect of various sparsification techniques on the main task and the attack accuracies.}
\label{tab:comp}
\end{table}

\paragraph{Compression. }
As detailed in Section \ref{sec:pre}, there could be a high communication cost in our application due to the vast number of trunk parameters. As these parameters are highly redundant, their lossy compression can significantly improve bandwidth efficiency at the cost of small performance degradation. Prior works also suggest that these techniques provide some privacy guarantees \cite{acs2012differentially,li2019privacy,zhou2009differential}.
As we show next, compression indeed mitigates the inference attacks described in Section \ref{sec:implemented_attacks}. Moreover, some sparsification techniques have a regularization effect on training, improving standard model accuracy.  

Compression techniques often rely on sparsifying the gradients (i.e., not transferring gradients with small magnitude), which can potentially mitigate inference attacks. Indeed, gradients modified by only a single training sample (revealing its membership) tend to have a smaller magnitude and are more likely to be eliminated by sparsification.

Three sparsification techniques are considered: \emph{thresholding}, \emph{Top-K} and \emph{random subset}. \emph{Thresholding} requires all participants to zero out all coordinate values of the update vector whose absolute value falls below a certain threshold and transfer only the remaining non-zero coordinates for aggregation. In contrast, via \emph{Top-K}, participants retain and transfer only the top $K$ largest coordinates in the update vector to the server. \emph{Random subset}, as its name suggests, transfers only a random subset of the gradients for aggregation. In particular, each participant sends the same random subset of all coordinates to the server, and they choose a new random subset in every round. This is implemented by sharing a common seed among all participants (since every participant has the same seed, they select the same coordinates in every round). Our results are presented in Table \ref{tab:comp}. Recall that an attack accuracy of 50\% indicates no information leakage concerning the considered attacks.

We found that a threshold of 0.01 keeps the number of non-zero gradients below 10\% in each training round and has excellent regularization properties. Indeed, thresholding increases model accuracy by 3\% compared to the baseline, uncompressed training. The right side of Figure \ref{fig:mit_NGMA} shows that, with Top-$K$, the less information we release about the gradients (i.e., the smaller $K$ is), the less effective the attack becomes.

\paragraph{Summary. }
We showed that simple hyper-parameter tuning, regularization, and compressing model updates could effectively reduce the feasibility of both membership attacks without degrading model accuracy. In particular, dropout, increasing the batch size, and applying thresholding on the model updates diminishes the feasibility of both attacks to ``Very difficult'', the likelihood of R1B to negligible without a significant drop in model accuracy. Moreover, besides eliminating both attacks, thresholding improves generalization and model accuracy. Differentiation attacks can be mitigated by mitigating membership attacks in the first place, by DP (at the cost of severe utility degradation), and by requiring fixed composition of clients during training. Moreover, as all clients participate in every federated round, the communication is synchronous, and all messages have the same size and format due to secure aggregation; hence T3 in Table \ref{tab:threaths} is also automatically mitigated since meta-data cannot be used to single-out any client during training. 

\section{Conclusions}\label{sec:conc}

%\begin{table}[t]
%    \centering
%    \begin{tabular}{|c|c|p{6cm}|}
%    \hline
%    \textbf{Feasibility} & \textbf{Level} & \textbf{Description} \\ \hline \hline
%    \cellcolor{green!25}Imp. & 1 & Almost impossible to perform. \\ \hline
%    \cellcolor{blue!25}Very Diff. & 2 & Difficult to perform even for an expert.\\ \hline
%    \cellcolor{yellow!25}Diff. & 3 & It is difficult to perform by average persons in the field. \\ \hline
%    \cellcolor{orange!25}Ave. & 4 & It can be done with some effort by an average person in the field. \\ \hline
%    \cellcolor{purple!25}Easy & 5 & Easy to perform by anyone in the field. \\ \hline
%    Immediate & 6 & Immediate feasibility by anyone. \\\hline
%    \end{tabular}
%    \caption{Feasibility}
%    \label{tab:feasibility}
%\end{table}

%\begin{table}[t]
%    \centering
%    \begin{tabular}{|M{.05\linewidth}|M{.4\linewidth}|M{.4\linewidth}|}
%    \hline
%    ID & Description & Mitigations \\
%    \hline 
%T1 &
%  Membership attack based on model updates, no direct assignment &
%  Regularization, parameter tuning, compression \\
%    \hline
%T2 &
%  Membership attack based on trunk activation, no direct assignment &
%  Regularization, compression \\
%    \hline
%T3 &
%  Differential attack based on trunk activation analysis with direct assignment &
%  Differential Privacy, Fixed composition of clients \\
%    \hline
%    \end{tabular}
%    \caption{Summary of identified risks and threats with their corresponding mitigation techniques. }
%    \label{tab:sum}
%\end{table}

The pharmaceutical industry can better leverage its data through collaboration. Yet, there are non-negligible risks stemming from the unintended information leakage, so such a platform needs to be secure and privacy-preserving. This paper described a privacy risk assessment of collaborative machine learning modeling in drug discovery. Several potential risks, threats, and attacks are discussed depending on the attacker's capability and goals. The main benefit of systematic risk assessment is prioritizing risks, which facilitates efficient mitigation of the most severe risks first.

Our novel attacks show that \emph{there is information leakage through the model updates}. Although there is limited leakage through the output of the commonly trained model, its feasibility is not always significant due to the strong assumption that the adversary needs access to labeled training data. We also emphasize that our attacks are evaluated on public data, and it may exhibit different performances on actual private pharma data. Moreover, we stress that membership attacks alone are insufficient to attribute a training sample to a specific participant. However, given a successful membership attack, such an attribution becomes feasible if the attacker can also launch a successful $N-1$ attack. This can happen when not all clients participate in every federated round, which can have many reasons such as client failure, DoS attack, the addition of new partners, etc. 

Besides analyzing the risks and threats, we described several mitigation techniques against them. Secure aggregation is an effective countermeasure as long as the composition of clients is fixed. Aside from threat mitigation, the proposed compression techniques also reduce the computational overhead of secure aggregation and communication costs in general. Differential privacy was also studied as it provides a provable defense against any attribution attack such as the $N-1$ attack. However, our experimental results suggest an intolerable accuracy drop, and thus we turned towards ad-hoc mitigation techniques, including regularization, hyper-parameter tuning, and compression of model updates.

We believe that our methodology can be generalized to other applications of federated multi-task learning. In particular, many models have sparse inputs with fully connected layers or are trained by a limited number of partners. For example, different health institutions can indirectly benefit from each other's patient databases through federated learning for different disease prognoses and predictions, especially if such diseases are rare. Therefore none of the partners alone has sufficient training data
\cite{healthfed}. Similarly, anomaly detection can also be significantly improved through federated learning due to the scarcity of anomalous patterns at a single organization or even in a smart building equipped with several IoT devices \cite{hamza}.

\paragraph{Acknowledgement.}
This work has received support from the EU/EFPIA Innovative Medicines Initiative 2 Joint Undertaking (MELLODDY grant nr 831472).
The research was supported by the Ministry of Innovation and Technology NRDI Office within the framework of the Artificial Intelligence National Laboratory Program.
The research reported in this paper is part of project no. BME-NVA-02, implemented with the support provided by the Ministry of Innovation and Technology of Hungary from the National Research, Development and Innovation Fund, financed under the TKP2021 funding scheme.
Project no. 138903 has been implemented with the support provided by the Ministry of Innovation and Technology from the NRDI Fund, financed under the FK 21 funding scheme.

%\paragraph*{Guidelines for Federated Learning Risk Mitigation}

%\begin{enumerate}
%    \item Use (Sparse) Secure Aggregation protocols to prevent the  trivial attribution of individual model updates (and ultimately sample membership information) to the participants, as well as to provide confidentiality to model updates.
%    \item Use membership attacks to test information leakage, if the formal privacy guarantee (e.g., differential privacy) is weak. Keep in mind that successful membership attack does not imply attribution, but only demonstrates information leakage.
%    \item Avoid changing the composition of participating parties during training to counter $N-1$ (differentiation) attack, and try to ensure that all parties participate in every single round of the training (e.g., mitigate DoS attacks and increase dependability in general).
%    \item Use hyper-parameter tuning, or add Gaussian noise to gradients to provide stronger formal privacy guarantee as well as to improve generalization. These techniques also increase robustness against attribution attacks.
%\end{enumerate}

\bibliographystyle{plain}
\bibliography{references.bib}

\end{document}